\documentclass[12pt]{article}

\catcode`\@=11
\global\arraycolsep=2pt
\usepackage{amsbsy,amssymb,latexsym,amsfonts,amsmath}
\usepackage{graphicx,color}

\begin{document}
\begin{flushright}
\parbox{4.2cm}
{}
\end{flushright}

\vspace*{0.7cm}

\begin{center}
{ \Large Hidden global conformal symmetry without Virasoro extension \\}
{ \Large  in theory of elasticity }
\vspace*{1.5cm}\\
{Yu Nakayama}
\end{center}
\vspace*{1.0cm}
\begin{center}

Kavli Institute for the Physics and Mathematics of the Universe (WPI),  
\\ University of Tokyo, 5-1-5 Kashiwanoha, Kashiwa, Chiba 277-8583, Japan

and
\\

Department of Physics, Rikkyo University, Toshima, Tokyo 175-8501, Japan

\vspace{3.8cm}
\end{center}

\begin{abstract}
The theory of elasticity (a.k.a. Riva-Cardy model) has been regarded as an example of scale invariant but not conformal field theories. We argue that in $d=2$ dimensions, the theory has hidden global conformal symmetry of $SL(2,\mathbb{R}) \times SL(2,\mathbb{R})$ without its Virasoro extension. More precisely,  we can embed all the correlation functions of the displacement vector into a global  conformal field theory with four-derivative action in terms of two scalar potential variables, which necessarily violates the reflection positivity.  The energy-momentum tensor for the potential variables cannot be improved to become traceless so that it does not show the Virasoro symmetry even with the existence of global special conformal current.

\end{abstract}

\thispagestyle{empty} 

\setcounter{page}{0}

\newpage

In recent years, the question under which condition scale invariance implies conformal invariance has gained some renewed interest (e.g. \cite{Nakayama:2010zz}\cite{Fortin:2012hn}\cite{Luty:2012ww}\cite{Bzowski:2014qja}\cite{Dymarsky:2014zja}\cite{Yonekura:2014tha}).\footnote{For a review and earlier reference, see \cite{Nakayama:2013is}.}
Due to the revival of conformal bootstrap in $d>2$ dimensions (see e.g. \cite{Qualls:2015qjb}\cite{Rychkov:2016iqz}\cite{Simmons-Duffin:2016gjk} and reference therein), the question has become not only academic but also practically important to reveal the nature of the renormalization group fixed point.

One interesting example of scale invariant but not conformal invariant field theories discussed in the literature is the theory of elasticity \cite{Landau}. For a displacement vector $v_\mu$, we have the (Euclidean) action 
\begin{align}
S = \int d^d x \left( -\frac{1}{4}(\partial_\mu v_\nu - \partial_\nu v_\mu)^2 - \frac{\kappa}{2} (\partial_\mu v^\mu)^2 \right) \ . 
\end{align}
Originally, Riva and Cardy \cite{Riva:2005gd} examined the model in $d=2$ dimensions. 
In general $d$-dimensions with generic $\kappa$, the theory has been regarded as an example of scale invariant but not conformal invariant field theories \cite{ElShowk:2011gz}.\footnote{In any dimensions, at $\kappa=1$, we have topologically twisted conformal symmetry. In $d > 2$ dimensions, we have untwisted (i.e. physical) conformal symmetry at $\kappa = (d-4)/d$.} Up to improvement terms, the symmetric energy-momentum tensor is given by
\begin{align}
T_{\mu\nu} &= \frac{\delta_{\mu\nu}}{4} (\partial_\rho v_\sigma - \partial_\rho v_\sigma)^2  -  (\partial_\mu v^\rho - \partial^\rho v_\mu) (\partial_\rho v_\nu - \partial_\nu v_\rho) \cr
&+ \kappa (v_\mu\partial_\nu (\partial^\rho v_\rho) + v_\nu \partial_\mu (\partial^\rho v_\rho) - \delta_{\mu\nu} (v_\rho\partial^\rho (\partial^\sigma v_\sigma) + (\partial^\rho v_\rho)^2/2)) \ .  
\end{align}

Let us recall that the condition of the scale invariance is that the trace of the symmetric energy-momentum tensor is given by divergence of the Virial current $J_\mu$ \cite{Wess}\cite{Coleman:1970je}.
\begin{align}
T^\mu_\mu = \partial^\mu J_\mu \ .
\end{align}
The condition for the global conformal invariance is that the Virial current is further rewritten as 
\begin{align}
J_\mu = \partial^\nu L_{\mu\nu} ,
\end{align}
with another symmetric tensor $L_{\mu\nu}$.
Indeed, if this is the case, we can construct a special conformal current
\begin{align}
J^{(k)}_\mu = T_{\mu\nu} ( 2k_\rho x^\rho x^\nu - x^2 k^\nu) - 2 k_\rho x^\rho \partial^\nu L_{\mu\nu} + 2k^\nu L_{\mu\nu} \label{specialconf}
\end{align}
that is conserved for a constant vector $k_\mu$ in any space-time dimensions (including $d=2$).

In $d=2$ dimensions, we typically observe that the global conformal symmetry of $SL(2,\mathbb{R}) \times SL(2,\mathbb{R})$ enjoys the Virasoro extension. The precise condition for the Virasoro extension is that the energy-momentum tensor is improved to become traceless. For this, we further need to require \cite{Polchinski:1987dy}\cite{Karananas:2015ioa}
\begin{align}
L_{\mu\nu} = \delta_{\mu\nu} L 
\end{align}
for a scalar operator $L$ so that we can construct the improved energy-momentum tensor that is traceless:
\begin{align}
\tilde{T}_{\mu\nu} &= T_{\mu\nu} + (\partial_\mu \partial_\nu - \delta_{\mu\nu}\Box ) L \ \cr
\tilde{T}^{\mu}_{\ \mu} &= 0
\end{align}
In $d=2$ dimensions, it then means that we  can construct the holomorphic energy-momentum tensor $T_{zz}(z)$ in the complex coordinate $z = x_1 + ix_2$, leading to the infinite number of conserved current $j^{(\epsilon)}(z) = \epsilon(z) T_{zz}(z)$ for any holomorphic function $\epsilon(z)$. 
Note that this condition is specific to the $d=2$ dimensions, and there is a niche possibility of having global conformal invariance without the Virasoro extension. As we will discuss at the end of this paper, the situation is rare, and it is only possible at the sacrifice of reflection positivity.

From the viewpoint of the Weyl invariance in the curved background, the special feature in $d=2$ dimension comes from the fact that the Ricci tensor and Ricci scalar are related as $R_{\mu\nu} = R g_{\mu\nu}/2$, so the curvature interaction $\int d^2x \sqrt{g} R_{\mu\nu} L^{\mu\nu}$ does not contain the symmetric traceless part. Therefore, in $d=2$ dimensions, the improvement term is able to remove the trace part of the symmetric tensor $L_{\mu\nu}$ but not the traceless part.

Let us go back to the theory of elasticity in $d=2$ dimensions. In $d=2$ dimensions, we can always (locally) decompose the vector field $v_\mu$ into potential variables  represented by a scalar $\phi$ and a pseudo-scalar $\chi$ as\footnote{It is known as the Helmholtz-Hodge decomposition. We assume that the space topology is trivial so that we have no global harmonic one-forms.}  
\begin{align}
v_\mu = \partial_\mu \phi + \epsilon_{\mu\nu} \partial^\nu \chi \ .
\end{align}
Substituting it into the action, we obtain
\begin{align}
S = \int d^2x \left( -\frac{1}{2} (\Box \chi)^2 -\frac{\kappa}{2} (\Box \phi)^2 \right) .
\end{align}
We now have two decoupled free scalar fields with four-derivative action (see \cite{Karananas:2015ioa} for a recent discussion on the model with its emphasis on the distinction between conformal invariance and Weyl invariance). 

Our claim is that if we use these potential variables, the theory of elasticity shows hidden global conformal invariance without the Virasoro extension for a generic value of $\kappa$.
To see the existence of global conformal symmetry first of all, let us compute the energy-momentum tensor in terms of $\phi$ and $\chi$ as
\begin{align}
T_{\mu\nu} &= -\delta_{\mu\nu}\left(\partial_\sigma \chi \partial^\sigma \Box \chi + \frac{1}{2}(\Box\chi)^2 \right) + \partial_\mu \Box \chi \partial_\nu \chi + \partial_\nu \Box \chi \partial_\mu \chi \cr
 & - \kappa \delta_{\mu\nu}\left(\partial_\sigma \phi \partial^\sigma \Box \phi + \frac{\kappa}{2}(\Box\phi)^2 \right) + \kappa \partial_\mu \Box \phi \partial_\nu \phi + \kappa \partial_\nu \Box \phi \partial_\mu \phi 
\end{align}
up to improvement terms (see e.g. \cite{Wiese:1996xd}\cite{Rajabpour:2011qr}\cite{Karananas:2015ioa}\cite{Osborn:2016bev}.
Its trace is 
\begin{align}
T_{\mu}^{\mu} = - (\Box \chi)^2 - \kappa (\Box \phi)^2 
\end{align}
so that it can be rewritten as
\begin{align}
T_{\mu}^{\mu} = \partial^\mu J_\mu = \partial^\mu \partial^\nu L_{\mu\nu} 
\end{align}
with
\begin{align}
L_{\mu\nu} =-2\partial_\mu \chi \partial_\nu \chi + \delta_{\mu\nu}\partial^\rho\chi \partial_\rho \chi + \delta_{\mu\nu} \chi \Box \chi
 +\kappa(-2\partial_\mu \phi \partial_\nu \phi + \delta_{\mu\nu}\partial^\rho\phi \partial_\rho \phi + \delta_{\mu\nu} \phi \Box \phi) \ 
\end{align}
up to the use of the equations of motion $\Box^2 \phi = \Box^2 \chi = 0$.

We, however, notice that $L_{\mu\nu} = \delta_{\mu\nu} L$ does not hold with a local operator $L$. From the above discussions, we see that the theory shows the global conformal symmetry\footnote{Since $\phi$ and $\chi$ decouple, the system actually possesses two copies of global conformal symmetry.} of $SL(2,\mathbb{R}) \times SL(2,\mathbb{R})$ but we cannot improve the energy-momentum tensor to become traceless. The theory does not exhibit the Virasoro symmetry (see also \cite{Rajabpour:2011qr}\cite{Karananas:2015ioa}).\footnote{We are informed that the case with $\kappa =0$ in the context of $d=2$ Maxwell theory was discussed in \cite{Ferrari:1992ic}\cite{Ferrari:1995gc}. We would like to thank S.~Rychkov for letting us know the above reference.}

More explicitly, let us use the complex coordinate $z = x_1 + ix_2$ and study the conformal transformation on the primary fields $\phi$ and $\chi$ as
\begin{align}
\delta \phi &= -\frac{1}{2} (\partial_z \epsilon)\phi + \epsilon \partial_z \phi \cr
\delta \chi & = -\frac{1}{2} (\partial_z \epsilon)\chi + \epsilon \partial_z \chi
\end{align}
 with respect to a holomorphic function $\epsilon(z)$. Up to the integration by part, the variation of the action is
\begin{align}
\delta_{\epsilon} S = \int d^2x (\partial_z^3 \epsilon) \left( -4(\bar{\partial}_{\bar{z}}\chi)^2 -4 \kappa  (\bar{\partial}_{\bar{z}}\phi)^2 \right)\ .
\end{align}
We see that the theory is invariant only under the global conformal transformation $\epsilon(z) = \epsilon_0 + \epsilon_1 z + \epsilon_2 z^2$, but not under the full conformal transformation in which $\epsilon(z)$ is an arbitrary holomorphic function.

The description in terms of the potential variable has a shift gauge symmetry
\begin{align}
\phi &\to \phi + c_{\phi} \cr
\chi & \to \chi + c_{\chi} \label{gauge}
\end{align}
with $c_{\phi}$ and $c_{\chi}$ being constant.
Therefore, all the correlation functions of $v_\mu$ in the original theory can be reproduced by those of the potential scalar fields $\phi$ and $\chi$, but not the other way. In particular, the potential variable $\phi$ and $\chi$ themselves cannot be expressed in terms of $v_\mu$ in a local manner. This is the reason why the global conformal symmetry is not manifest in terms of $v_\mu$. Indeed, the special conformal current \eqref{specialconf} is not invariant under the shift gauge symmetry \eqref{gauge}.

More abstractly, all the correlation functions of $v_\mu$ can be obtained from the Wick contraction with
\begin{align}
\langle v_\mu(p) v_\nu (q) \rangle = \delta(p+ q) \frac{1}{p^2} \left(\delta_{\mu\nu} - (1-\kappa^{-1}) \frac{p_\mu p_\nu}{p^2} \right) \ 
\end{align}
in the momentum space.
In terms of the potential variables, they can be reproduced by $v_{\mu} =  \partial_\mu \phi + \epsilon_{\mu\nu} \partial^\nu \chi$ and using the Wick contraction of the potential scalars with
\begin{align}
\langle \phi(p) \phi(q) \rangle &= \delta(p+q) \frac{\kappa^{-1}}{p^4} \cr
\langle \chi (p) \chi(q)\rangle &= \delta(p+q) \frac{1}{p^4} \ .
\end{align}
These scalar two-point functions are consistent with global conformal invariance (but not with the Virasoro symmetry) in which $\phi$ and $\chi$ are (quasi-)primary fields of the dimensions $\Delta(\chi) = \Delta(\phi) = -1$. However, $\phi$ and $\chi$ themselves do not appear in the correlation functions of $v_\mu$. Thus, more strictly speaking, we may declare that all the correlation functions of $v_\mu$ in the theory of elasticity can be embedded into a global conformal invariant field theory realized by free scalars $\phi$ and $\chi$. See \cite{ElShowk:2011gz}\cite{Dymarsky:2015jia} for a similar concept of embedding of scale invariant field theories into conformal field theories.

As it turns out, the situation is reminiscent of the three-dimensional free Maxwell theory
\begin{align}
 S = \int d^3x \left(\frac{1}{4} F_{\mu\nu} F^{\mu\nu} \right) \ ,
\end{align}
in which the  theory does not show manifest conformal symmetry in terms of the electric description based on the vector potential $A_\mu$, but shows hidden conformal symmetry in terms of the magnetic variable  $B$ with $\partial_\mu B = \epsilon_{\mu\nu\rho} F^{\nu\rho}$ with the dual action \cite{Jackiw:2011vz}\cite{ElShowk:2011gz}
\begin{align}
S = \int d^3x  \left(\frac{1}{2}\partial_\mu B \partial^\mu B \right) \ .
\end{align}
Again, there is a constant shift gauge symmetry $B \to B +c$, but the special conformal current in terms of $B$ is not invariant under this gauge symmetry.

In this paper, we have shown that the theory of elasticity in $d=2$ dimensions has  hidden global conformal symmetry $SL(2,\mathbb{R}) \times SL(2,\mathbb{R})$ without its Virasoro extension. A lesson we can draw from this is that a possibility of (infinite dimensional) extension of the algebra does not necessarily imply the physical realization of such extended algebra. Indeed, there are many non-relativistic examples of (infinite dimensional) extended algebra whose physical realization seem impossible \cite{Soojong}.
To conclude the paper we have two final remarks.

Firstly, the peculiarity that we have observed is possible only because the theory does not possess the reflection positivity. The canonical scaling of the energy-momentum tensor implies the symmetric traceless tensor $L_{\mu\nu}$ has scaling dimension zero in $d=2$ dimensions, and it is inconsistent with the reflection positivity. As discussed in \cite{Riva:2005gd}, the theory of elasticity lacks the reflection positivity.

Secondly, the similar models of scale invariant vector theories appear in the theory of perception \cite{Bialek:1986it}\cite{Bialek:1987qc}\cite{Nakayama:2010ye}. There, the approximate symmetry of our two-dimensional perception (i.e. our visual perception of two-dimensional image) given by Euclidean invariance and scale invariance was associated with the manifest symmetry of the underlying field theory. Our analysis, however, shows that they may possess the hidden global conformal symmetry in addition. It would be interesting to understand its role in the theory of perception. Perhaps we may enhance our visual perception by exploiting the hidden symmetry that we have discussed in this paper.

\section*{Acknowledgements}
The author would like to thank T.~Okuda for asking the question if there is a possibility of having global conformal invariance without Virasoro symmetry several years ago, at which time the author gave the wrong answer. 
This work was in part supported by the World Premier International Research Center Initiative (WPI Initiative), MEXT.


\begin{thebibliography}{99}

%\cite{Nakayama:2010zz}
\bibitem{Nakayama:2010zz} 
  Y.~Nakayama,
  %``Scale invariance vs conformal invariance from holography,''
  Int.\ J.\ Mod.\ Phys.\ A {\bf 25}, 4849 (2010).
  doi:10.1142/S0217751X10050731
  %%CITATION = doi:10.1142/S0217751X10050731;%%
  %16 citations counted in INSPIRE as of 23 Mar 2016


%\cite{Luty:2012ww}
\bibitem{Luty:2012ww} 
  M.~A.~Luty, J.~Polchinski and R.~Rattazzi,
  %``The $a$-theorem and the Asymptotics of 4D Quantum Field Theory,''
  JHEP {\bf 1301}, 152 (2013)
  doi:10.1007/JHEP01(2013)152
  [arXiv:1204.5221 [hep-th]].
  %%CITATION = doi:10.1007/JHEP01(2013)152;%%
  %111 citations counted in INSPIRE as of 23 Mar 2016



%\cite{Fortin:2012hn}
\bibitem{Fortin:2012hn} 
  J.~F.~Fortin, B.~Grinstein and A.~Stergiou,
  %``Limit Cycles and Conformal Invariance,''
  JHEP {\bf 1301}, 184 (2013)
  doi:10.1007/JHEP01(2013)184
  [arXiv:1208.3674 [hep-th]].
  %%CITATION = doi:10.1007/JHEP01(2013)184;%%
  %54 citations counted in INSPIRE as of 23 Mar 2016

%\cite{Bzowski:2014qja}
\bibitem{Bzowski:2014qja} 
  A.~Bzowski and K.~Skenderis,
  %``Comments on scale and conformal invariance,''
  JHEP {\bf 1408}, 027 (2014)
  doi:10.1007/JHEP08(2014)027
  [arXiv:1402.3208 [hep-th]].
  %%CITATION = doi:10.1007/JHEP08(2014)027;%%
  %12 citations counted in INSPIRE as of 23 Mar 2016

%\cite{Dymarsky:2014zja}
\bibitem{Dymarsky:2014zja} 
  A.~Dymarsky, K.~Farnsworth, Z.~Komargodski, M.~A.~Luty and V.~Prilepina,
  %``Scale Invariance, Conformality, and Generalized Free Fields,''
  JHEP {\bf 1602}, 099 (2016)
  doi:10.1007/JHEP02(2016)099
  [arXiv:1402.6322 [hep-th]].
  %%CITATION = doi:10.1007/JHEP02(2016)099;%%
  %20 citations counted in INSPIRE as of 23 Mar 2016

%\cite{Yonekura:2014tha}
\bibitem{Yonekura:2014tha} 
  K.~Yonekura,
  %``Unitarity, Locality, and Scale versus Conformal Invariance in Four Dimensions,''
  arXiv:1403.4939 [hep-th].
  %%CITATION = ARXIV:1403.4939;%%
  %2 citations counted in INSPIRE as of 23 Mar 2016



%\cite{Nakayama:2013is}
\bibitem{Nakayama:2013is} 
  Y.~Nakayama,
  %``Scale invariance vs conformal invariance,''
  Phys.\ Rept.\  {\bf 569}, 1 (2015)
  doi:10.1016/j.physrep.2014.12.003
  [arXiv:1302.0884 [hep-th]].
  %%CITATION = doi:10.1016/j.physrep.2014.12.003;%%
  %70 citations counted in INSPIRE as of 23 Mar 2016


%\cite{Qualls:2015qjb}
\bibitem{Qualls:2015qjb} 
  J.~D.~Qualls,
  %``Lectures on Conformal Field Theory,''
  arXiv:1511.04074 [hep-th].
  %%CITATION = ARXIV:1511.04074;%%
  %3 citations counted in INSPIRE as of 25 Mar 2016

%\cite{Rychkov:2016iqz}
\bibitem{Rychkov:2016iqz} 
  S.~Rychkov,
  %``EPFL Lectures on Conformal Field Theory in $D \ge 3$ Dimensions,''
  arXiv:1601.05000 [hep-th].
  %%CITATION = ARXIV:1601.05000;%%
  %8 citations counted in INSPIRE as of 25 Mar 2016

%\cite{Simmons-Duffin:2016gjk}
\bibitem{Simmons-Duffin:2016gjk} 
  D.~Simmons-Duffin,
  %``TASI Lectures on the Conformal Bootstrap,''
  arXiv:1602.07982 [hep-th].
  %%CITATION = ARXIV:1602.07982;%%
  %1 citations counted in INSPIRE as of 23 Mar 2016

\bibitem{Landau}
 L.~D.~Landau and E.~M.~Lifshitz, Theory of elasticity, Pergamon,
New York, 1970.

%\cite{Riva:2005gd}
\bibitem{Riva:2005gd} 
  V.~Riva and J.~L.~Cardy,
  %``Scale and conformal invariance in field theory: A Physical counterexample,''
  Phys.\ Lett.\ B {\bf 622}, 339 (2005)
  doi:10.1016/j.physletb.2005.07.010
  [hep-th/0504197].
  %%CITATION = doi:10.1016/j.physletb.2005.07.010;%%
  %46 citations counted in INSPIRE as of 23 Mar 2016

%\cite{ElShowk:2011gz}
\bibitem{ElShowk:2011gz} 
  S.~El-Showk, Y.~Nakayama and S.~Rychkov,
  %``What Maxwell Theory in D<>4 teaches us about scale and conformal invariance,''
  Nucl.\ Phys.\ B {\bf 848}, 578 (2011)
  doi:10.1016/j.nuclphysb.2011.03.008
  [arXiv:1101.5385 [hep-th]].
  %%CITATION = doi:10.1016/j.nuclphysb.2011.03.008;%%
  %57 citations counted in INSPIRE as of 23 Mar 2016

\bibitem{Wess}
J.~Wess, ``The Conformal Invariance in Quantum Field Theory", Nuovo Cim. 18 (1960) 1086.

\bibitem{Coleman:1970je}
  S.~R.~Coleman and R.~Jackiw,
  ``Why dilatation generators do not generate dilatations?,''
  Annals Phys.\  {\bf 67} (1971) 552.
  %%CITATION = APNYA,67,552;%%


%\cite{Polchinski:1987dy}
\bibitem{Polchinski:1987dy} 
  J.~Polchinski,
  %``Scale and Conformal Invariance in Quantum Field Theory,''
  Nucl.\ Phys.\ B {\bf 303}, 226 (1988).
  doi:10.1016/0550-3213(88)90179-4
  %%CITATION = doi:10.1016/0550-3213(88)90179-4;%%
  %229 citations counted in INSPIRE as of 23 Mar 2016




%\cite{Karananas:2015ioa}
\bibitem{Karananas:2015ioa} 
  G.~K.~Karananas and A.~Monin,
  %``Weyl vs. Conformal,''
  arXiv:1510.08042 [hep-th].
  %%CITATION = ARXIV:1510.08042;%%
%\cite{Wiese:1996xd}
\bibitem{Wiese:1996xd} 
  K.~J.~Wiese,
  %``Classification of perturbations for membranes with bending rigidity,''
  Phys.\ Lett.\ B {\bf 387}, 57 (1996)
  doi:10.1016/0370-2693(96)00989-6
  [cond-mat/9607192].
  %%CITATION = doi:10.1016/0370-2693(96)00989-6;%%
  %2 citations counted in INSPIRE as of 16 Apr 2016
%\cite{Rajabpour:2011qr}
\bibitem{Rajabpour:2011qr} 
  M.~A.~Rajabpour,
  %``Conformal symmetry in non-local field theories,''
  JHEP {\bf 1106}, 076 (2011)
  doi:10.1007/JHEP06(2011)076
  [arXiv:1103.3625 [hep-th]].
  %%CITATION = doi:10.1007/JHEP06(2011)076;%%
  %3 citations counted in INSPIRE as of 16 Apr 2016


%\cite{Osborn:2016bev}
\bibitem{Osborn:2016bev} 
  H.~Osborn and A.~Stergiou,
  %``$C_T$ for Non-unitary CFTs in Higher Dimensions,''
  arXiv:1603.07307 [hep-th].
  %%CITATION = ARXIV:1603.07307;%%




%\cite{Ferrari:1992ic}
\bibitem{Ferrari:1992ic} 
  F.~Ferrari,
  %``Free and interacting 2-D Maxwell field theory on conformally flat space-times,''
  Class.\ Quant.\ Grav.\  {\bf 10}, 1065 (1993)
  doi:10.1088/0264-9381/10/6/005
  [hep-th/9211045].
  %%CITATION = doi:10.1088/0264-9381/10/6/005;%%
  %6 citations counted in INSPIRE as of 16 Apr 2016

%\cite{Ferrari:1995gc}
\bibitem{Ferrari:1995gc} 
  F.~Ferrari,
  %``Biharmonic conformal field theories,''
  Phys.\ Lett.\ B {\bf 382}, 349 (1996)
  doi:10.1016/0370-2693(96)00677-6
  [hep-th/9507142].
  %%CITATION = doi:10.1016/0370-2693(96)00677-6;%%
  %1 citations counted in INSPIRE as of 16 Apr 2016


%\cite{Dymarsky:2015jia}
\bibitem{Dymarsky:2015jia} 
  A.~Dymarsky and A.~Zhiboedov,
  %``Scale-invariant breaking of conformal symmetry,''
  J.\ Phys.\ A {\bf 48}, no. 41, 41FT01 (2015)
  doi:10.1088/1751-8113/48/41/41FT01
  [arXiv:1505.01152 [hep-th]].
  %%CITATION = doi:10.1088/1751-8113/48/41/41FT01;%%
  %5 citations counted in INSPIRE as of 23 Mar 2016

%\cite{Jackiw:2011vz}
\bibitem{Jackiw:2011vz} 
  R.~Jackiw and S.-Y.~Pi,
  %``Tutorial on Scale and Conformal Symmetries in Diverse Dimensions,''
  J.\ Phys.\ A {\bf 44}, 223001 (2011)
  doi:10.1088/1751-8113/44/22/223001
  [arXiv:1101.4886 [math-ph]].
  %%CITATION = doi:10.1088/1751-8113/44/22/223001;%%
  %41 citations counted in INSPIRE as of 23 Mar 2016


%\cite{Bialek:1986it}
\bibitem{Bialek:1986it} 
  W.~Bialek and A.~Zee,
  %``Statistical Mechanics and Invariant Perception,''
  Phys.\ Rev.\ Lett.\  {\bf 58}, 741 (1987).
  doi:10.1103/PhysRevLett.58.741
  %%CITATION = doi:10.1103/PhysRevLett.58.741;%%
  %9 citations counted in INSPIRE as of 23 Mar 2016

%\cite{Bialek:1987qc}
\bibitem{Bialek:1987qc} 
  W.~Bialek and A.~Zee,
  %``Understanding the Efficiency of Human Perception,''
  Phys.\ Rev.\ Lett.\  {\bf 61}, 1512 (1988).
  doi:10.1103/PhysRevLett.61.1512
  %%CITATION = doi:10.1103/PhysRevLett.61.1512;%%
  %6 citations counted in INSPIRE as of 23 Mar 2016

%\cite{Nakayama:2010ye}
\bibitem{Nakayama:2010ye} 
  Y.~Nakayama,
  %``Gravity Dual for a Model of Perception,''
  Annals Phys.\  {\bf 326}, 2 (2011)
  doi:10.1016/j.aop.2010.09.009
  [arXiv:1003.5729 [hep-th]].
  %%CITATION = doi:10.1016/j.aop.2010.09.009;%%
  %6 citations counted in INSPIRE as of 23 Mar 2016

\bibitem{Soojong}
Y.~Nakayama and S.~J.~Rey, unpublished.

\end{thebibliography}
\end{document}